\documentclass{elsart} 
\usepackage{epsfig,amssymb,amsmath,cite}
\journal{Annals of Physics}
\date{24 May 2006}
\begin{document}

\begin{frontmatter}

\title{Exactly solvable associated Lam\'e potentials and supersymmetric transformations}

\author[Mexico]{David J. Fern\'andez C.\corauthref{cor}},
\corauth[cor]{Corresponding author.} \ead{david@fis.cinvestav.mx}
\author[Kolkata,Valladolid]{Asish Ganguly}
\ead{gangulyasish@rediffmail.com}
\address[Mexico]{Departamento de F\'{\i}sica, Cinvestav, AP 14-740, 07000
M\'exico DF, Mexico}
\address[Kolkata]{City College (C.C.C.B.A.), University of Calcutta,
13 Surya Sen Street, Kolkata 700 012, India}
\address[Valladolid]{Departamento de F\'{\i}sica
Te\'{o}rica, At\'{o}mica y \'{O}ptica, Universidad de Valladolid,
47071 Valladolid, Spain}
Accepted in Annals of Physics
\begin{abstract}
A systematic procedure to derive exact solutions of the associated
Lam\'e equation for an arbitrary value of the energy is presented.
Supersymmetric transformations in which the seed solutions have
factorization energies inside the gaps are used to generate new
exactly solvable potentials; some of them exhibit an interesting
property of periodicity defects.
\end{abstract}

\begin{keyword}
Supersymmetric quantum mechanics \sep associated Lam\'e potentials
\PACS 11.30.Pb \sep 03.65.Ge \sep 03.65.Fd\sep 02.30.Gp
\end{keyword}

\end{frontmatter}


\section{Introduction}

The exactly solvable potentials of the one-dimensional Schr\"odinger
equation are important due to the fact that the full physical
information of the system is encoded in a small number of analytical
expressions. Moreover, they can be used to test the convergence of
the numerical methods as well as to be the departure point for
applying the widely used perturbative techniques. By exact
solvability we mean that the stationary Schr\"odinger equation for
the involved potential (when it is periodic) admits analytic
solutions for energies in the allowed bands as well as for  energies
in the gaps.

A remarkable fact, the existence of a class of potentials which are
intermediate between the exactly solvable ones and those just
solvable through numerical techniques, was realized since the date
back to 80´s of past century. Nowadays they are known as
quasi-exactly solvable potentials, and their characteristic property
is that there exist compact analytical expressions of physical
information for only a part of the system spectrum. This means that
there are some missing features of these models which just can be
numerically determined. A big effort has been observed for years in
identifying the potentials which are quasi-exactly solvable, and
gradually it started to dominate the conviction that this class
includes the associated Lam\'e potentials
\cite{ks99,ga00,ga02,ds02,tv02,rkp04,fg05,al05}, among others.
However, some signs have been recently noticed indicating that the
associated Lam\'e potentials for some integer values of the
parameter pair $(m,\ell)$ belong as well to the exactly solvable
class. Recently, we proposed \cite{fg05} an ansatz through which one
can implement a fitting procedure to automatically fix the values of
$m, \ \ell$, the ansatz parameters and consequently the analytic
solutions of the associated Lam\'e equation. Unfortunately, that
procedure was not completely systematic in the sense that one has to
compute case by case for each pair of values of $(m,\ell)$ to find
the general solution by adopting the natural modification of the
ansatz.

On the other hand, it seems interesting to enlarge the exactly
solvable class of potentials departing from a given initial one.
Several procedures are available to do this, and the simplest one is
called supersymmetric quantum mechanics (SUSY QM)
\cite{ba00,afhnns04,mr04,nnr04,ac04,io04,su05,ff05}. In this
approach there is a differential operator of order $k$ intertwining
the initial and final Hamiltonians, the main ingredients being $k$
seed solutions of the initial Schr\"odinger equation. These seeds
can be chosen either physical (as it was done previously when using
the ground state) or non-physical. When the last one are used, it
has been possible to surpass the usual restriction of the standard
first-order SUSY QM that the new levels will be created below the
ground state energy of the initial Hamiltonian \cite{ff05}.

The supersymmetric transformations started to be implemented just
few year ago to periodic potentials
\cite{df98,ks99,fnn00,nnf00,fmrs02a,fmrs02b,ro03,bgbm99,gt04}.
Similarly as for the non-periodic case, at the beginning physical
seed solutions were used (band edge eigenfunctions)
\cite{df98,ks99,fnn00,nnf00}. However, very soon it was realized
that non-physical seed solutions could be as well employed. In this
way, it was possible to generate either periodic potentials (when
non-physical Bloch solutions were used) or non-periodic ones (with
periodicity defects) when general linear combinations of the
non-physical Bloch solutions were employed
\cite{fmrs02a,fmrs02b,ro03,fg05}.

For the associated Lam\'e potentials the supersymmetric
transformations were applied quite recently \cite{fg05}. However,
the treatment was done for first-order transformations and for
particular values of the parameter pair $(m, \ell)$. It would be
important to implement the SUSY QM for $k>1$ and for general integer
values of the parameter pair $(m, \ell)$.

The main aim of this paper is to present a systematic procedure to
derive general solutions of the one-dimensional Schr\"odinger
equation for the associated Lam\'e potential with an arbitrary
energy $E$
\begin{eqnarray}
& H \psi(x) =
\left[-\partial^2_x+V(\!x)\right]\psi(x)=E\psi(x), \nonumber \\
& V(x) = m(m+1)k^2{\rm sn}^2x+\ell(\ell+1)k^2\frac{{\rm cn}^2x}{{\rm
dn}^2x}, \label{aj}
\end{eqnarray}
which will work, in principle, for any integer values of the
parameters $m$ and $\ell$. Our technique is based on the well known
Frobenius method, and we will naturally arrive at two separate
cases, characterized either by $m>\ell$ or by $m=\ell$. A
fundamental conclusion of our treatment is that the associated
Lam\'e potentials for any integer values of the parameters $m$ and
$\ell$ are exactly solvable.

In addition, by implementing the supersymmetric transformations (by
choosing the obtained non-physical solutions) we will generate new
exactly solvable potentials from the associated Lam\'e equation. We
will restrict ourselves, by simplicity, to first and second-order
transformations, but this procedure can be continued at will for any
order of the intertwining operator.

Next section will be devoted on the discussion of our procedure at
length to derive the general solutions of the associated Lam\'e
equation (for readers' convenience a brief introduction about
elliptic functions is included in the Appendix). We will apply, in
the subsequent sections the supersymmetric transformations, of first
and higher order, to generate new exactly solvable potentials, which
can have periodic structure or periodicity defects depending on how
we choose the initial seed Schr\"odinger solutions. We will end the
paper with our conclusion.

\section{General solution of associated Lam\'e equation}

Let us make some preliminary remarks about the equation (\ref{aj}).
It is defined on the full real line ($x\in \mathbb{R})$ and the
modulus parameter $k^2$ of Jacobian elliptic functions belongs to
the interval $(0,1)$. The potential is non-singular and periodic of
real period $2K$ or $K$ according to $m\neq \ell$ or $m=\ell$
respectively, where
$K(k)=\int_0^{\pi/2}d\phi/\sqrt{1-k^2\sin^2\phi}$.  Also it is
sufficient to consider $m\ge \ell$ for their non-negative integer
values (see the discussions in Sec III, Ref. \cite{ga02}). The
equation (\ref{aj}) may be transformed by applying a coordinate
translation
\begin{equation}\label{tr1}
x\rightarrow z=\frac{x-iK'}{\sqrt{\bar{e}_{3}}}, \quad K'\equiv
K(k'), \quad k'^2=1-k^2,
\end{equation}
to Weierstrass form
\begin{equation}
\label{aw} -\frac{d^2\psi}{dz^2}+ \left[m(m+1)\wp(z)+
\frac{\ell(\ell+1)\bar{e}_2\bar{e}_3}{\wp(z)-e_1}\right]\psi=\widetilde{E}\psi,
\quad \psi(x)\equiv \psi (z(x)),
\end{equation}
where
\begin{equation}\label{def1}
\widetilde{E}=e_3m(m+1)+[E-\ell(\ell+1)]\bar{e}_3, \qquad
\bar{e}_i=e_1-e_i, \ i=2,3.
\end{equation}

In equation (\ref{aw}) $\wp(z)\equiv \wp(z;\omega,\omega')$ is
Weierstrass elliptic function of half-periods
$\omega=K/\sqrt{\bar{e}_3},\ \omega'=iK'/\sqrt{\bar{e}_3}$ and the
real numbers $e_i \ (e_1>e_2>e_3)$ are related with Weierstrass
invariants through the definition $\wp(\omega_i)=e_i, \
\omega_1\equiv \omega, \ \omega_2\equiv \omega+\omega', \
\omega_3=\omega'$. Let us denote two linearly independent solutions
of equation (\ref{aw}) by $\psi^+(z),\psi^-(z)$\footnote{For
convenience we use this notation to distinguish two solutions and
should not be confused with the popular notation for intertwined
eigenstates.}. Then their product $\Psi(z)=\psi^+\psi^-$ will be a
solution of the following third order differential equation
\begin{displaymath}
\frac{d^3\Psi}{dz^3}-4\left[m(m+1)\wp(z)+\frac{\ell(\ell+1)
\bar{e}_2\bar{e}_3}{\wp(z)-e_1}- \widetilde{E}\right]
  \frac{d\Psi}{dz}
\end{displaymath}
\begin{equation}
\label{prod1}\hspace{1cm}-2\left[m(m+1)-\frac{\ell(\ell+1)\bar{e}_2\bar{e}_3}{[
\wp(z)-e_1]^2}\right]\wp'(z)\Psi=0,
\end{equation}
where, throughout this article the prime will denote differentiation
with respect to its argument. It may be mentioned that in the limit
$\ell=0$ or $-1$ Schr\"odinger equation (\ref{aj}) reduces to the
well-known Lam\'e equation and consequently the singularity of
(\ref{prod1}) at $z=\omega_1$ disappears, i.e., the singularities
remain only at the poles of $\wp(z)$ in the complex $z$-plane. We
will now express equation (\ref{prod1}) in algebraic form by
applying the transformations
\begin{equation}\label{tr2}
y=\frac{e_1-\wp(z)}{\bar{e}_2}, \qquad
\Phi(y)=[\wp(z)-e_1]^\ell\Psi.
\end{equation}
The equation (\ref{prod1}) then reduces to (see the Appendix for
more details)
\begin{equation}
\label{prod2}P_4(y)\frac{d^3\Phi}{dy^3}+P_3(y)\frac{d^2\Phi}{dy^2}+P_2(y)\frac{d\Phi}{dy}+P_1(y)\Phi=0,
\end{equation}
in which $P_i(y)$ are $i$-th degree polynomials in $y$ given by
\begin{eqnarray}
&& P_4(y)\!=2y^2(\bar{e}_2y^2-\!3e_1y+\bar{e}_3),
\hspace{7.5cm}\hspace{-.2915pt} \nonumber \\
&&
P_3(y)\!=3y[\bar{e}_2(3\!-\!2\ell)y^2\!-\!6e_1(1\!-\!\ell)y+\!\bar{e}_3(1\!-\!2\ell)],
\nonumber \\
&& P_2(y)=2\{\bar{e}_2(3\ell^2-m^2-6\ell-m+3)y^2 \nonumber \\
&& \hspace{1.3cm}-[\widetilde{E}+e_1(9\ell^2-m^2-9\ell-m+3)]y
          +\ell(2\ell-1)\bar{e}_3\},\nonumber \\
&& P_1(y)=\bar{e}_2(2\ell-1)(m+\ell)(m-\ell+1)y+2\ell[\widetilde{E}
+e_1(3\ell^2-m^2-m)]. \nonumber
\end{eqnarray}

The differential equation (\ref{prod2}) is clearly of Fuschian type
having four regular singular points at $y=0,1,\bar{e}_3/\bar{e}_2$
and $\infty$. It is evident that one can construct by Frobenius
method, in general, a power series solution around any singular
point which will be valid in a circle of convergence containing no
other singularity. Such solutions are therefore of local nature. We
are interested in global solutions, if exist, valid in the whole
$y$-plane. But this means that three local solutions around three
singular points in the finite part must coincide, which is possible
iff the Frobenius series terminates after a finite number of terms.
However, a formal solution around $y=0$ can be considered in the
form
\begin{equation}
\label{fs}\Phi=\sum_{r=0}^\infty a_ry^{r+\rho}.
\end{equation}
The indicial equation and the recurrence relations for the
coefficients are obtained in straightforward way as follows
\begin{eqnarray}
\label{ie} & a_0 f_0(\rho)=0, \qquad a_1f_0(\rho+1)+a_0f_1(\rho)=0,
\\ \label{re}
&\hskip-0.8cm
a_{r+2}f_0(\rho+r+2)+a_{r+1}f_1(\rho+r+1)+a_rf_2(\rho+r)=0, \ \ \
r=0,1,2 \dots ,
\end{eqnarray}
where
\begin{eqnarray}\label{def}
\label{f0}f_0(\rho) & = & \bar{e}_3\rho(\rho-1-2\ell)(2\rho-2\ell-1), \\
\label{f1}f_1(\rho) & = & 2(\rho-\ell)\{e_1[m(m+1)-3(\rho-\ell)^2]-\widetilde{E}\},\\
\label{f2}f_2(\rho) & = &
\bar{e}_2(\rho-m-\ell)(\rho+m-\ell+1)(2\rho+1-2\ell).
\end{eqnarray}

From the recurrence relation (\ref{re}), it is clear that one can
not get a finite series corresponding to $\rho=2\ell+1$ and
$\ell+1/2$. The smallest exponent $0$ is, in general,
non-preferable, since it differs from the greatest exponent by an
integer and so leads to the solution involving logarithmic terms.
But it is known that under certain conditions \cite{in56}
logarithmic terms may not appear in the leading solution. We will
now investigate this possibility. Note that $f_0(2\ell+1)=0$ and so
for $r=2\ell-1$, recurrence relation (\ref{re}) reduces to a
two-term relation for $\rho=0$
\begin{equation}
\label{rre}a_{2\ell}f_1(2\ell)+a_{2\ell-1}f_2(2\ell-1)=0
\end{equation}
and $a_{2\ell+1}$ remains arbitrary up to now. However, the
coefficients $a_r$ for $r<2\ell+1$ could be determined explicitly in
the form
\begin{equation}
\label{a2l}a_r=(-1)^r a_0 F_r/\prod^r_{s=1}f_0(s) \, ,\quad r=1,2,
                       \dots 2\ell,
\end{equation}
where $F_r$ is an $r\times r$ determinant :
\begin{equation}
\label{Fr} F_r=\left |
  \begin{array}{llllll}
     f_1(r-1)       &      f_2(r-2)        &   \hspace{.7cm} 0    &   \hspace{.7cm} 0   & \cdots & \hspace{.3cm} 0 \\
     f_0(r-1)       &      f_1(r-2)        &       f_2(r-3)       &   \hspace{.7cm} 0   & \cdots & \hspace{.3cm} 0 \\
  \hspace{.7cm} 0   &      f_0(r-2)        &       f_1(r-3)       &      f_2(r-4)       & \cdots & \hspace{.3cm} 0 \\
\hspace{.4cm} \cdots & \hspace{.4cm} \cdots & \hspace{.4cm} \cdots & \hspace{.4cm}\cdots & \cdots &    \cdots       \\
  \hspace{.7cm} 0   &   \hspace{.7cm} 0    &   \hspace{.7cm} 0    &   \hspace{.7cm} 0   & \cdots &    f_1(0)       \\
      \end{array} \right |
\end{equation}

The first step is to show that $F_{2\ell+1}\equiv 0$, which will
ensure the absence of logarithmic terms from the solution. Close
inspection of equations (\ref{f0}-\ref{f2}) shows the following
interesting properties of $f_0,f_1,f_2$ for $\nu=1,2,3,\ldots$
\begin{equation}
\label{fprop}
f_0(2\ell-\nu)=-f_0(\nu+1),\:f_1(2\ell-\nu)=-f_1(\nu),\:
f_2(2\ell-\nu-1)=-f_2(\nu).
\end{equation}
It follows at once from equation (\ref{Fr}) and (\ref{fprop}) that
$F_{2\ell+1}$ is skew-symmetric and so it vanishes identically.
Next, we observe that $F_{2\ell+1}$ may be expressed as
$F_{2\ell+1}=f_1(2\ell)F_{2\ell}-f_0(2\ell)f_2(2\ell-1)F_{2\ell-1}$,
which proves the consistency of the constraint relation (\ref{rre}).

Our final step is to fix $a_{2\ell+1}$ in such a way that, if
possible, the series (\ref{fs}) for $\rho=0$ terminates, say, after
$(N+1)$ terms. This is equivalent to impose the following additional
relations
\begin{equation}
\label{are1}a_Nf_1(N)+a_{N-1}f_2(N-1)=0, \qquad a_{N+1}=0,
\end{equation}
\begin{equation}
\label{are2}f_2(N)=0,
\end{equation}
where the integer $N$ is to be determined. The equation (\ref{are2})
clearly gives acceptable value of $N$ for
\begin{equation}\label{nfix}
N=m+\ell.
\end{equation}
But then (\ref{are1}) reads
\begin{equation}
\label{fre} a_{m+\ell}f_1(m+\ell)+a_{m+\ell-1}f_2(m+\ell-1)=0,
\qquad a_{m+\ell+1}=0.
\end{equation}
We now consider two cases separately to check consistency of
relation (\ref{fre}) and, if consistent, to fix
$a_{2\ell+1},a_{2\ell+2},\ldots a_{m+\ell}$.

\vspace{.5cm} \noindent
\mbox{\textbf{case i)$\qquad m=\ell$}}\\

\noindent In this case relation (\ref{fre}) coincides with
(\ref{rre}) and so it is clearly consistent. Further equation
(\ref{fre}) fixes $a_{2\ell+1}=0$. As a result, the series
(\ref{fs}) for $\rho=0$ terminates after $(2\ell+1)$ terms.

\vspace{.5cm} \noindent
\mbox{\textbf{case ii)$\qquad m=\ell+\nu,\: \nu=1,2, \ldots$}}\\

\noindent In this case recurrence relations (\ref{re}) and
(\ref{fre}) give $\nu$ relations for $\nu$ unknowns
$a_{2\ell+1},a_{2\ell+2},\ldots a_{2\ell+\nu}$
\begin{eqnarray*}
& a_{2\ell+2}f_0(2\ell+2)+
a_{2\ell+1}f_1(2\ell+1)+a_{2\ell}f_2(2\ell)= 0, \\
& a_{2\ell+3}f_0(2\ell+3)+ a_{2\ell+2}f_1(2\ell+2)+
a_{2\ell+1}f_2(2\ell+1)= 0, \\
& \vdots
\\
& a_{2\ell+\nu}f_0(2\ell+\nu)+ a_{2\ell+\nu-1}f_1(2\ell+
\nu-1)+a_{2\ell+ \nu-2}f_2(2\ell+\nu-2) = 0, \\
& a_{2\ell+\nu}f_1(2\ell+\nu) + a_{2\ell+\nu-1}f_2(2\ell+\nu-1) = 0.
\end{eqnarray*}
Then by back substitution we find
\begin{equation}
\label{fa2l} a_{2\ell+r} = \frac{(-1)^rD_{\nu-r}
\prod_{s=0}^{r-1}f_2(2\ell+s)}{D_{\nu}}a_{2\ell}, \quad r=1,2,\ldots
\nu,
\end{equation}
where $D_r$ is the minor of $F_{2\ell+\nu+1-r}$ in Laplace expansion
of the determinant $F_{2\ell+\nu+1}$. This means that the $r\times
r$ determinant $D_r$ is obtained from $F_{2\ell+\nu+1}$ by
suppressing $(2\ell+\nu+1-r)$ rows and columns in which
$F_{2\ell+\nu+1-r}$ is placed. Thus we get a finite series
(\ref{fs}) terminating after $(2\ell+\nu)$ terms, since
$a_{2\ell+\nu+1}=a_{2\ell+\nu+2}=\cdots=0$.

Hence, we have proved that the differential equation (\ref{prod2})
possesses a polynomial solution of degree $(m+\ell)$ of the form
\begin{equation}\label{prs}
\Phi=\sum^{m+\ell}_{r=0}a_ry^r, \qquad a_0\neq 0,
\end{equation}
for a special choice of the $a_r$'s. The coefficients $a_r$ for
$r>2\ell$ may be expressed in a compact form, valid for both cases
$m=\ell$ and $m>\ell$
\begin{equation}
\label{cfa2l} a_{2\ell+r} = \left (
(-1)^rD_{m-\ell-r}\prod^{r-1}_{s=0}f_2(2\ell+s)/D_{m-\ell} \right
) a_{2\ell}, \: r=1,2,\ldots ; \  D_{-r}\equiv 0.
\end{equation}
We recall that the $a_r$ for $r\leq 2\ell$ are given by equation
(\ref{a2l}), while the rest for $r>2\ell$ are to be determined from
(\ref{cfa2l}). This means that all the coefficients in the series
(\ref{prs}) can be expressed in terms of the normalization constant
$a_0$, which may be taken as 1. Thus, the product $\Psi(z)$ of two
linearly independent solutions $\psi^+(z),\psi^-(z)$ of the
associated Lam\'e equation (\ref{aw}) may be written in the form (up
to some inessential constant factor)
\begin{equation}
\label{pp}\Psi(z)=
\frac{\prod_{r=1}^{m+\ell}\left[\wp(z)-\wp(b_r)\right]}{[\wp(z)-e_1]^\ell},
\end{equation}
where $\wp(b_1),\wp(b_2),\ldots \wp(b_{m+\ell})$ are the zeros of
the polynomial $\sum_{r=0}^{m+\ell}a_r[(e_1-t)/\bar{e}_2]^r$, $a_r$
being determined from (\ref{a2l}) and (\ref{cfa2l}). At this moment
it is worth mentioning that for practical computation of $b_r$ one
has to invert the transcendental relation $\wp(b_r)=c_r$, and so,
due to the fact that $\wp(z)$ is an even function, an ambiguity of
sign appears in the process, which we shall fix now. Let us first
express the two linearly independent solutions $\psi^+(z),\psi^-(z)$
in terms of their product $\Psi(z)$. Note that the Wronskian
$W(\psi^+,\psi^-)$ must be non-vanishing and without loss of
generality we may set
\begin{displaymath}
W\equiv 1=\psi^+(\psi^-)'-(\psi^+)'\psi^-.
\end{displaymath}
Dividing the above identity by $\Psi=\psi^+\psi^-$,
\begin{equation}\label{wr0}
(\ln \psi^-)'-(\ln \psi^+)'=\frac{1}{\Psi}.
\end{equation}

Performing the logarithmic differentiation of the product solution
$\Psi=\psi^+\psi^-$ with respect to $z$, we arrive
\begin{equation}
\label{wr} (\ln \psi^-)'+(\ln \psi^+)'=(\ln \Psi)'.
\end{equation}
Thus, adding and subtracting the relations (\ref{wr0}) and
(\ref{wr}),
\begin{equation}
\label{ln1}(\ln \psi^-)'=\frac{1}{2}\left[(\ln
\Psi)'+\frac{1}{\Psi}\right], \qquad (\ln
\psi^+)'=\frac{1}{2}\left[(\ln \Psi)'-\frac{1}{\Psi}\right].
\end{equation}
From (\ref{ln1}) it follows readily
\begin{equation}
\label{usp}\psi^{\pm}(z)=\sqrt{\Psi(z)}\exp\left
(\mp\frac{1}{2}\int^z\frac{d\tau}{\Psi(\tau)}\right ).
\end{equation}

To fix the sign of $b_r$, we will now differentiate the second
relation in  (\ref{ln1}):
\begin{displaymath}
\frac{(\psi^+)''}{\psi^+}-\left[\frac{(\psi^+)'}{\psi^+}\right]^2=\frac{1}{2}\left[\frac{\Psi''}{\Psi}
-\left(\frac{\Psi'}{\Psi}\right)^2+\frac{\Psi'}{\Psi^2}\right].
\end{displaymath}
Substitution for $(\psi^+)'/\psi^+\equiv(\ln\psi^+)'$ from
(\ref{ln1})
\begin{equation}\label{aux}
 \frac{(\psi^+)''}{\psi^+}=\frac{1}{4\Psi^2}\left(2\Psi\Psi'' +
 1-\Psi'^2\right).
\end{equation}
Now multiplying equation (\ref{aux}) by $4\,\Psi^2$, and using the
associated Lam\'e equation (\ref{aw}) for the solution $\psi^+$, we
obtain
\begin{equation}\label{fix}
2\Psi\Psi'' + 1-\Psi'^2 =
4\Psi^2\left[m(m+1)\wp(z)+\frac{\ell(\ell+1)
\bar{e}_2\bar{e}_3}{\wp(z)-e_1}-\widetilde{E}\right].
\end{equation}
Noting that $b_r$ are the zeros of $\Psi(z)$ [see equation
(\ref{pp})], for the values $z=b_r, r=1,2,\ldots m+\ell$, equation
(\ref{fix}) gives
\begin{equation}
\label{fix1} \left . \Psi^{'^{2}}\right |_{z=b_r}=1.
\end{equation}
The ambiguity of signs in the values of $b_r$ may now be fixed by
selecting the convention
\begin{equation}\label{ff}
\Psi'|_{z=b_r}=+1,
\end{equation}
which can also be expressed in terms of $\wp$-functions by
evaluating $\Psi'$ at $z=b_j$ from (\ref{pp}) as
\begin{equation}\label{ffp}
\Psi'|_{z=b_j}\equiv
\frac{\wp'(b_j)}{[\wp(b_j)-e_1]^{\ell}}\prod^{m+
\ell}_{\renewcommand{\arraystretch}{.3}
                       \begin{array}{c}r=1 \\
                                    r\neq j
                       \end{array}} \renewcommand{\arraystretch}{1}
                                  \left[\wp(b_j)-\wp(b_r)\right]=+1, \quad j=1,2,\ldots m+\ell.
\end{equation}

The remaining job now is to express the integrand in (\ref{usp}) in
a form suitable for performing the integration. Let us express
$1/\Psi$, with $\Psi$ given by (\ref{pp}), as a sum of partial
fractions in the form
\begin{displaymath}
\frac{1}{\Psi(z)}=\sum^{m+\ell}_{r=1}\frac{A_r}{\wp(z)-\wp(b_r)}.
\end{displaymath}
Then, equating each term from both sides of the above identity for
$z=b_1,b_2,\ldots b_{m+\ell}$, and using the relation (\ref{ffp}),
it is straightforward to obtain
\begin{equation}
\label{prec}\frac{1}{\Psi(z)}=\sum_{r=1}^{m+\ell}\frac{\wp'(b_r)}{\wp(z)-\wp(b_r)}.
\end{equation}
To proceed further we introduce two quasi-periodic functions
$\zeta(z)$ and $\sigma(z)$ by the definitions
\begin{equation}\label{defz}
\zeta'(z)=-\wp(z), \qquad [\ln\sigma(z)]'=\zeta(z),
\end{equation}
with the properties
\begin{equation}\label{prop1}
\zeta(z+2\omega_i)=\zeta(z)+2\zeta(\omega_i), \qquad
\zeta(-z)=-\zeta(z),
\end{equation}
\begin{equation}\label{prop2}
\sigma(z+2\omega_i)= -\exp[2\zeta(\omega_i)(z+\omega_i)]\sigma(z),
            \quad \sigma(-z)=-\sigma(z).
\end{equation}
These are known as Weierstrass zeta and sigma functions
respectively. The addition formulae for them are
\begin{equation}\label{ad1}
\zeta(z+y)=\zeta(z)+\zeta(y)+\frac{1}{2}\frac{\wp'(z)-\wp'(y)}{\wp(z)-\wp(y)},
\end{equation}
\begin{equation}\label{ad2}
 \sigma(z+y)\sigma(z-y)=-\sigma^2(z)\sigma^2(y)[\wp(z)-\wp(y)].
\end{equation}

For our purpose it is useful to rewrite the addition formula
(\ref{ad1}) in the form
\begin{equation}\label{ad1f}
\zeta(z-y)-\zeta(z+y)=\frac{\wp'(z)}{\wp(z)-\wp(y)}-2\zeta(y),
\end{equation}
where we have used the fact that $\wp(z)$ is an even function, but
$\wp'(z), \ \zeta(z)$ and $\sigma(z)$ are odd. It is now not very
difficult to rewrite the equation (\ref{prec}) by replacing $y$ in
(\ref{ad1f}) by $b_r$ :
\begin{equation}
\label{frec}\frac{1}{\Psi(z)}=\sum_{r=1}^{m+\ell}[\zeta(z-b_r)-\zeta(z+b_r)+2\zeta(b_r)].
\end{equation}
The advantage of writing $1/\Psi$ in the form (\ref{frec}) is that
one can express the argument of the exponential term in the solution
(\ref{usp}) in a compact form by exploiting the definitions
(\ref{defz})
\begin{equation}
\label{int}\int^z\frac{d\tau}{\Psi(\tau)}=\sum_{r=1}^{m+\ell}\left [
\ln\frac{\sigma(z-b_r)}{\sigma(z+b_r)}+2z\zeta(b_r)\right ].
\end{equation}

The factor $\sqrt{\Psi(z)}$ in the solution (\ref{usp}) may also be
expressed in terms of sigma and zeta functions by using (\ref{pp})
and (\ref{ad2}) as (up to some factor)
\begin{equation}\label{pp1}
 \sqrt{\Psi(z)}=\frac{\exp [\ell z\zeta
 (\omega_1)]}{\sigma^m(z)\sigma^{\ell}(z+\omega_1)}\prod^{m+\ell}_{r=1}
 \sqrt{\sigma(z+b_r)\sigma(z-b_r)}.
\end{equation}
It may be mentioned that in the above equation we have eliminated
the term $\sigma(z-\omega_1)$ by using the quasi-periodic property
(\ref{prop2}). Our solutions of the associated Lam\'e equation
(\ref{aj}) for any integral values of $m,\ell$ will now be
presented in their final form
\begin{equation}
\label{fins}
\psi^{\pm}(x)=\frac{\prod_{r=1}^{m+\ell}\sigma(\frac{x-iK'}{\sqrt{\bar{e}_3}}\pm
b_r)}{\sigma^{\ell}(\frac{x-iK'}{\sqrt{\bar{e}_3}}+\omega_1)\sigma^{m}(\frac{x-iK'}{\sqrt{\bar{e}_3}})}
\exp\left\{\frac{x}{\sqrt{\bar{e}_3}}\left[\ell \zeta(\omega_1)\mp
\sum_{r=1}^{m+\ell}\zeta(b_r)\right]\right\}.
\end{equation}

It is instructive to consider the limit $\ell\rightarrow 0$, which
will reduce the associated Lam\'e equation (\ref{aj}) to the
ordinary Lam\'e equation
\begin{equation}\label{lame}
 [-\partial_x^2+m(m+1)k^2{\rm sn}^2x]\psi(x)=E\psi(x).
\end{equation}
The solution of (\ref{lame}) can be obtained from (\ref{fins}) in
the limit $\ell\rightarrow 0$
\begin{equation}\label{ls}
\psi^{\pm}(x)=\frac{\prod_{r=1}^{m}\sigma(\frac{x-iK'}{\sqrt{\bar{e}_3}}\pm
b_r)}{\sigma^{m}(\frac{x-iK'}{\sqrt{\bar{e}_3}})} \exp
\left[\mp\frac{x}{\sqrt{\bar{e}_3}} \sum_{r=1}^{m}\zeta(b_r)\right],
\end{equation}
which is exactly identical with the old result obtained for Lam\'e
equation in Ref. \cite{ww63}. Further, it may also be noted that the
general solution (\ref{fins}) perfectly coincides with our previous
results \cite{fg05} for $(m,\ell)=(1,1)$ and $(2,1)$. In the next
section we will apply the supersymmetric transformations using
generalized superpotentials based on the general solution
(\ref{fins}).

\section{Supersymmetric transformations}

A simple technique to generate new Hamiltonians $\widetilde H$ with
known spectra from a given initial one $H$ is the so-called
supersymmetric quantum mechanics (SUSY QM)
\cite{ba00,afhnns04,mr04,nnr04,ac04,io04,su05,ff05}. In this
procedure the non-null action of a finite-order differential
intertwining operator $B$ such that
\begin{equation}
\widetilde H B = B H
\end{equation}
onto the eigenfunctions of $H$ provides those of $\widetilde H$
(which in particular is valid for the physical eigenfunctions). In
the modern approach to the subject the intertwining operator $B$ is
of $k$-th order, and its simplest variant involves $k$ seed
Schr\"odinger solutions of $H$ associated to $k$ different
factorization energies, namely,
\begin{equation}
H u_i = \epsilon_i u_i,
\end{equation}
which are annihilated as well by $B$, i.e.,
\begin{equation}
B u_i = 0, \ i=1,\dots,k.
\end{equation}
The new potential $\widetilde V(x)$ differs from the initial one
$V(x)$ by a term involving the Wronskian $W(u_1,\dots,u_k)$ of the
$k$ seed solutions in the way
\begin{equation}
\widetilde V(x)=V(x) - 2[\ln W(u_1,\dots,u_k)]''.
\end{equation}
It is worth to notice that this generalized approach to SUSY QM
allows to surpass the typical restriction of the standard
first-order version that the new levels have to be created below the
ground state energy $E_0$ of $H$. This is possible since the SUSY
transformations can be implemented using either physical
Schr\"odinger solutions (as it was typically done, e.g., through the
use of the ground state) or non-physical ones \cite{fg05}.

The SUSY QM was applied for a long time to Hamiltonians with a
discrete part of the spectrum (see e.g. the collection of articles
in \cite{afhnns04}). Except by few previous works \cite{tr89},
however, the SUSY transformations started to be implemented in a
systematic way just recently to potentials which are periodic,
specifically to the Lam\'e potentials of (\ref{lame})
\cite{df98,ks99,fnn00,nnf00,fmrs02a,fmrs02b,ro03,bgbm99,gt04}.
Similarly as for the non-periodic case, for the Lam\'e equation
(\ref{lame}) the SUSY transformations were implemented first by
using seed physical solutions (the band edge eigenfunctions of $H$)
\cite{df98,ks99,fnn00,nnf00}. However, very soon it was understood
that more general eigenfunctions of $H$ (non-physical ones included)
provide once again a generalized scheme \cite{fmrs02a,fmrs02b,ro03}.
In particular, it includes the possibility of generating either
periodic or non-periodic  (with periodicity defects) SUSY partner
potentials, which depends on the use of Bloch solutions (\ref{ls})
or their general linear combinations for a fixed set of
factorization energies.

Concerning the associated Lam\'e potentials (\ref{aj}), the SUSY QM
has been applied recently \cite{fg05}. The treatment was restricted
to first-order transformations, the parameter pair $(m,\ell)$ taking
the values $(1,1)$ and $(2,1)$. In this paper we will implement the
supersymmetric transformations to the associated Lam\'e potentials
for any integer value of the pair $(m,\ell)$. For the sake of
simplicity, we will restrict ourselves to first and second-order
transformations.

\subsection{First-order SUSY QM}

Let us apply then the SUSY QM of first order to the associated
Lam\'e potentials. We denote by $u(x)$ the involved seed
Schr\"odinger solution and by $\epsilon$ the corresponding
factorization energy. In order to avoid the singularities in
$\widetilde V(x)$ we will suppose that $\epsilon \leq E_0$. This
means that we need to know the exact information of the ground state
in order to choose $\epsilon$ appropriately. Fortunately this
information is already available in the literature
\cite{ks99,ga00,ga02}.

We will use in the first place any of the two Bloch solutions
(\ref{fins}) as transformations functions, i,e., we take $u(x) =
\psi^{\pm}(x)$. Therefore:
\begin{eqnarray}
[\ln u(x)]'' & = & \frac{m}{{\rm sn}^2(x -i K')} + \ell +
\frac{\ell(2 e_1^2 + e_2 e_3)}{\bar e_3^2}\frac{{\rm sn}^2(x -i
K')}{{\rm cn}^2(x - i K')} \nonumber \\ && \hskip2cm -
\sum_{r=1}^{m+\ell} \frac{1}{{\rm sn}^2(x \pm \sqrt{\bar e_3}\,b_r-i
K')}.
\end{eqnarray}
By employing that $k^2{\rm sn}^2(x -i K') = 1/{\rm sn}^2x$ and
$k^2{\rm cn}^2(x - i K') = -{\rm dn}^2x /{\rm sn}^2x$ one arrives
to:
\begin{equation}
[\ln u(x)]'' = m k^2 {\rm sn}^2x + \ell - \frac{\ell(2 e_1^2 + e_2
e_3)}{\bar e_3^2{\rm dn}^2x} - k^2 \sum_{r=1}^{m+\ell} {\rm sn}^2(x
\pm \sqrt{\bar e_3}b_r). \label{deltav1}
\end{equation}
Thus, the periodic first-order SUSY partner potentials generated
through the Bloch solutions (\ref{fins}) become:
\begin{equation}
\widetilde V_\pm(x) = m(m-1)k^2{\rm sn}^2x + \ell (\ell - 1)k^2
\frac{ {\rm cn}^2x}{{\rm dn}^2x} + 2 k^2 \sum_{r=1}^{m+\ell} {\rm
sn}^2(x \pm \sqrt{\bar e_3}b_r). \label{tvp1}
\end{equation}
Since the intertwining operator $B$ transforms bounded Schr\"odinger
solutions into bounded ones, unbounded into unbounded, etc., it
turns out that the new potentials have exactly the same band
spectrum as the associated Lam\'e potentials.

On the other hand, if we use as seed a nodeless linear combination
of the two Bloch solutions (\ref{fins}) associated to a given
factorization energy $\epsilon \leq E_0$,
\begin{equation}
u(x) = A \psi^+(x) + B \psi^-(x) = A \psi^+(x)\phi_+(x) = B
\psi^-(x)\phi_-(x), \label{linearcombination}
\end{equation}
where $\lambda_+ = B/A, \ \lambda_- = A/B$, and
\begin{equation}
\phi_\pm(x) = 1 + \lambda_\pm \frac{\psi_{\mp}(x)}{\psi_{\pm}(x)} =
1 + \lambda_\pm
\prod_{r=1}^{m+\ell}\frac{\sigma(\frac{x-iK'}{\sqrt{\bar{e}_3}}\mp
b_r)}{\sigma(\frac{x-iK'}{\sqrt{\bar{e}_3}}\pm b_r)}
\exp\left[\frac{2x\zeta(b_r)}{\sqrt{\bar{e}_3}}\right],
\end{equation}
it turns out that the SUSY partners for the associated Lam\'e
potentials become now:
\begin{equation}
\widetilde V^{np}(x) = \widetilde V_\pm (x) - 2 [\ln \phi_\pm(x)]'',
\label{tvnp1}
\end{equation}
where $\widetilde V_\pm (x)$ are given by (\ref{tvp1}).

Let us notice that $\widetilde V^{np}(x)$ is non-periodic in the
full real line, a behavior characterized specifically by the term
$[\ln \phi_\pm(x)]''$. In fact, there is a periodicity defect in a
finite region of $x$, but the potential acquires an asymptotic
periodic behavior when we move far away of that domain. Let us
remark that in the direct approach, even if we would know how to do
it, it would be difficult to determine the spectrum of the
Hamiltonians associated to the non-periodic potentials $\widetilde
V^{np}(x)$. However, due to its periodic behavior for large $\vert
x\vert$ and since the intertwining operator maps bounded
Schr\"odinger solutions into bounded ones, unbounded into unbounded,
etc., it turns out that the SUSY approach provides in a simple way
the spectra of the new Hamiltonians, which contain the allowed
energy bands of the initial associated Lam\'e potential. Moreover,
it can be shown that $1/u(x)$ is square-integrable, meaning that
$\widetilde V^{np}(x)$ acquires an extra isolated bound state at
$\epsilon$.

\subsection{Second-order SUSY QM}

Let us apply now the second-order SUSY QM by using two Bloch
solutions of the form (\ref{fins}) associated to two different
factorization energies $\epsilon_1, \epsilon_2$, where we denote by
$b_r$ and $b_r'$ the corresponding constants. In order to avoid the
zeros in the Wronskian, which would produce singularities in
$\widetilde V(x)$, we choose $\epsilon_1, \epsilon_2$ to be in the
same forbidden gap of $H$ (for a concrete proof see \cite{fmrs02a});
this includes of course the possibility for both to be in the
infinity gap below $E_0$ but it allows as well to use solutions in
finite gaps (above $E_0$). Once again, this means that we will need
information about the positions of the band edges, which will be
taken from earlier works \cite{ks99,ga00,ga02}. We will select, for
definiteness both solutions with the upper signs; any other signs
combination will lead essentially to the same expressions which will
be derived below. Thus we take
\begin{eqnarray}
&& \hskip-1cm u_{1}(x)=\psi_1^+(x) =
\frac{\prod_{r=1}^{m+\ell}\sigma(\frac{x-iK'}{\sqrt{\bar{e}_3}}+
b_r)}{\sigma^{\ell}(\frac{x-iK'}{\sqrt{\bar{e}_3}}+\omega_1)\sigma^{m}(\frac{x-iK'}{\sqrt{\bar{e}_3}})}
\exp\big\{\frac{x}{\sqrt{\bar{e}_3}}\big[\ell \zeta(\omega_1)-
\sum_{r=1}^{m+\ell}\zeta(b_r)\big]\big\} \nonumber \\
&& \hskip-1cm u_{2}(x)= \psi_2^+(x) =
\frac{\prod_{r=1}^{m+\ell}\sigma(\frac{x-iK'}{\sqrt{\bar{e}_3}}+
b'_r)}{\sigma^{\ell}(\frac{x-iK'}{\sqrt{\bar{e}_3}}+\omega_1)\sigma^{m}(\frac{x-iK'}{\sqrt{\bar{e}_3}})}
\exp\big\{\frac{x}{\sqrt{\bar{e}_3}}\big[\ell \zeta(\omega_1)-
\sum_{r=1}^{m+\ell}\zeta(b'_r)\big]\big\}
\end{eqnarray}
An appropriate expression for the Wronskian reads:
\begin{equation}
W(u_1,u_2) =  g(x) \, u_1(x) \, u_2(x), \quad g(x) =
\left[\ln\left(\frac{u_2}{u_1}\right) \right]',
\end{equation}
where, by using the previous formulae for $u_1(x), \ u_2(x)$, it can
be shown that:
\begin{eqnarray}
g(x) = &  \frac{1}{\sqrt{\bar e_3}}
\sum_{r=1}^{m+\ell}\big[\zeta\big(\frac{x-iK'}{\sqrt{\bar{e}_3}}+
b'_r\big)- \zeta(b'_r) - \zeta\big(\frac{x-iK'}{\sqrt{\bar{e}_3}}+
b_r\big)+ \zeta(b_r)\big] \label{f}
\end{eqnarray}
Thus, the modification to the potential is given by:
\begin{eqnarray}
\left[\ln W(u_1,u_2) \right]'' & = &  \left(\ln g \right)'' +
\left(\ln u_1  \right)'' + \left(\ln u_2 \right)''\nonumber
\\ & = & \left(\ln g \right)'' + 2 m k^2 {\rm sn}^2x + 2\ell - 2\frac{\ell(2 e_1^2 + e_2
e_3)}{\bar e_3^2{\rm dn}^2x} \nonumber \\
&& - k^2\sum_{r=1}^{m+\ell} \left[{\rm sn}^2(x + \sqrt{\bar e_3}b_r)
+ {\rm sn}^2(x + \sqrt{\bar e_3}b'_r)\right],
\end{eqnarray}
where we have used (\ref{deltav1}) to simplify this expression.
Finally, the periodic second-order SUSY partner potential
$\widetilde V(x)$ of $V(x)$ is given by:
\begin{eqnarray}
\widetilde V(x) & = & m(m-3)k^2{\rm sn}^2x + \ell (\ell -
3)k^2 \frac{{\rm cn}^2x}{{\rm dn}^2x} \nonumber \\
&& + 2 k^2\sum_{r=1}^{m+\ell}\left[{\rm sn}^2(x + \sqrt{\bar
e_3}b_r)+ {\rm sn}^2(x + \sqrt{\bar e_3}b'_r)\right] - 2\left(\ln g
\right)''. \label{tvp}
\end{eqnarray}

On the other hand, if we use as seeds two general linear
combinations of the Bloch solutions associated to $\epsilon_1, \
\epsilon_2$, which up to unimportant constant factors can be
expressed as
\begin{equation}\label{lc12}
u_1(x)=\psi_1^+ + \lambda_1 \psi_1^- = \psi_1^+ \phi_1^+, \qquad
     u_2(x)=\psi_2^+ + \lambda_2\psi_2^- = \psi_2^+ \phi_2^+,
\end{equation}
where
\begin{eqnarray}
\phi_1^+ & = & 1 + \lambda_1 \frac{\psi_1^-}{\psi_1^+} = 1 +
\lambda_1 \prod_{r=1}^{m+\ell}
\frac{\sigma(\frac{x-iK'}{\sqrt{\bar{e}_3}}-
b_r)}{\sigma(\frac{x-iK'}{\sqrt{\bar{e}_3}}+ b_r)} \exp\left[\frac{2
x\zeta(b_r)}{\sqrt{\bar{e}_3}}\right], \nonumber \\
\phi_2^+ & = & 1 + \lambda_2 \frac{\psi_2^-}{\psi_2^+} = 1 +
\lambda_2 \prod_{r=1}^{m+\ell}
\frac{\sigma(\frac{x-iK'}{\sqrt{\bar{e}_3}}-
b'_r)}{\sigma(\frac{x-iK'}{\sqrt{\bar{e}_3}}+ b'_r)}
\exp\left[\frac{2 x\zeta(b'_r)}{\sqrt{\bar{e}_3}}\right],
\end{eqnarray}
we will arrive at non-periodic second-order SUSY partner potentials.
Indeed, a convenient expression for the Wronskian reads:
\begin{equation}
W(u_1,u_2) = u_1 u_2 g^{np} = \psi_1^+ \psi_2^+ \phi_1^+ \phi_2^+
g^{np},
\end{equation}
where
\begin{equation}
g^{np} = \left[\ln\left(\frac{u_2}{u_1}\right) \right]' =
\left[\ln\left(\frac{\psi_2^+}{\psi_1^+}\right) \right]'+
\left[\ln\left(\frac{\phi_2^+}{\phi_1^+}\right) \right]' = g +
\left[\ln\left(\frac{\phi_2^+}{\phi_1^+}\right) \right]',
\end{equation}
with $g$ given by (\ref{f}). The modification of the potential is
thus given by:
\begin{equation}
\left[\ln W(u_1,u_2) \right]'' = \left[\ln\left( g \psi_1^+ \psi_2^+
\right) \right]'' + \left[\ln \left(\phi_1^+ \phi_2^+
\frac{g^{np}}{g}\right) \right]''.
\end{equation}
With this equation it is simple to see that the non-periodic
second-order SUSY partner of the associated Lam\'e potential
becomes:
\begin{equation}
\widetilde V^{np}(x) =  \widetilde V(x) - 2 \left[\ln\left(\phi_1^+
\phi_2^+ \frac{g^{np}}{g}\right) \right]'', \label{tvnp2}
\end{equation}
where $\widetilde V(x)$ is given by (\ref{tvp}). Let us notice that
this part of $\widetilde V^{np}(x)$ coincides with the periodic
second-order SUSY partner potential previously derived. The
non-periodicity in $\widetilde V^{np}(x)$ is produced by the second
term in the RHS of equation (\ref{tvnp2}), but now this term is more
involved than in the first-order case. However, the general behavior
of $\widetilde V^{np}(x)$ is quite similar, i.e., there is a finite
region in the $x$-domain in which a periodicity defect is produced.
Hence, similar to the case for first-order SUSY we have obtained a
potential, which is asymptotically periodic. It is worth to mention
that the spectrum of $\widetilde V^{np}(x)$ contains again the
allowed energy bands of the initial associated Lam\'e potential, but
now there will be two extra bound states at $\epsilon_1, \
\epsilon_2$ which could be interesting for physical applications.

\begin{figure}[ht]
\centering \epsfig{file=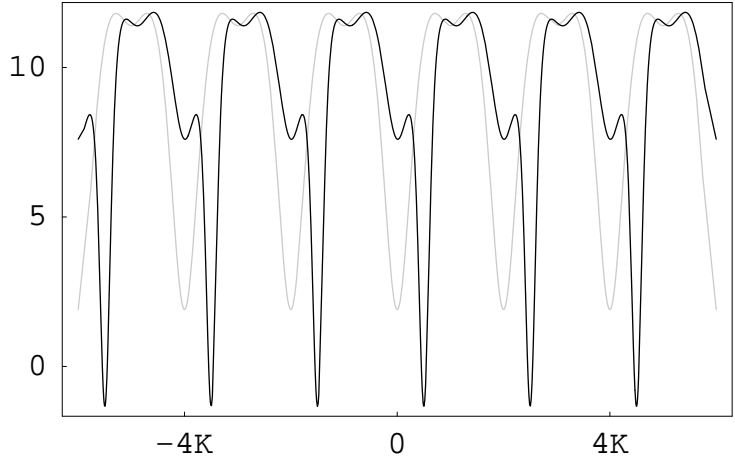, width=10cm} \caption{\small
Periodic first-order SUSY partner potential $\widetilde V_+(x)$
(black curve) isospectral to the associated Lam\'e potential (gray
curve) with $m=3, \ \ell=1$, $k^2=0.95$, generated by using the
Bloch solution $u(x) = \psi^+(x)$ with $\epsilon = 4.75$.}
\end{figure}

\section{Applications}

In a previous paper \cite{fg05} we have found solutions of the
stationary Schr\"odinger equation with an arbitrary energy (either
physical or non-physical) for the associated Lam\'e potentials, the
pair $(m,\ell)$ taking the values $(1,1)$ and $(2,1)$. We applied
there also the first-order SUSY techniques in order to generate new
potentials with known spectra. Here, we will illustrate our previous
general results with a different case characterized by
$(m,\ell)=(3,1)$. For this associated Lam\'e potential explicit
expressions for some band-edge eigenfunctions and eigenvalues are
known \cite{ga02}\footnote{We will denote here by $\psi_{(i)}$ the
band edge eigenfunctions to distinguish them from the solutions used
in the SUSY transformations. Some mistakes in the ordering of levels
in Ref. \cite{ga02} have been corrected here.}:
\begin{eqnarray*}
&& \psi_{(1)} = {\rm cn}\,x \, {\rm dn}^2x, \hskip2cm E_1 = 1 + 4 k^2, \nonumber \\
&& \psi_{(2)} = {\rm sn}\,x \, {\rm dn}^2x, \hskip2cm E_2 = 1 + 9
k^2, \nonumber
\end{eqnarray*}
\vskip-1cm
\begin{eqnarray*}
\psi_{(3,8)} & = & \frac{{\rm sn}\,x \, {\rm cn}\,x}{{\rm
dn}\,x}\left[{\rm
sn}^2x - \frac{1}{5k^2}\left(k^2 + 3\pm\sqrt{k^4-9k'^2}\right)\right], \nonumber \\
E_{3,8} & = & 10 + 2 k^2 \mp 2 \sqrt{k^4-9k'^2}.
\end{eqnarray*}
Three other band edge eigenfunctions can be given:
\begin{eqnarray*}
&\hskip-1cm \psi_{(i)} = \frac{{\rm sn}^4x}{{\rm dn}\,x} - \frac{(9
k^2+16-E_i)}{10k^2} \frac{{\rm sn}^2x}{{\rm dn}\,x} +
\frac{E_i^2-2(5k^2+18)E_i+ 9k^4+156k^2+320}{15k^4{\rm dn}\,x}
\end{eqnarray*}
where the eigenvalues $E_i, \ i = 0,4,7$ are the ordered roots of
the cubic equation:
\begin{equation*}
E^3-(11k^2+20)E^2+(19k^4+216k^2+64)E-(9k^6+388k^4+448k^2)=0
\end{equation*}

\begin{figure}[ht]
\centering \epsfig{file=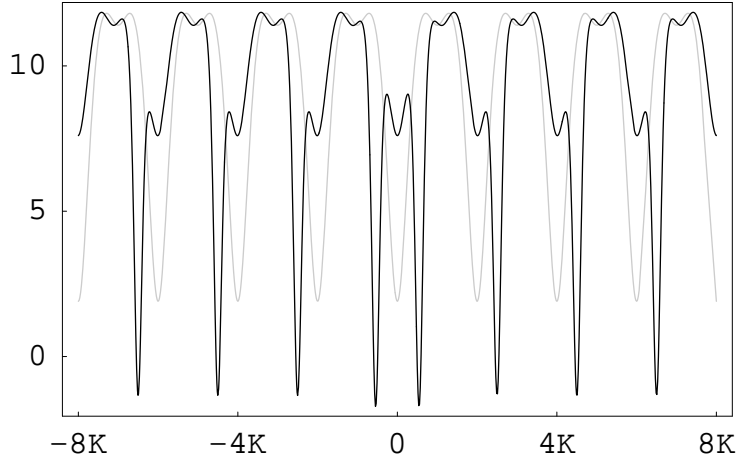, width=10cm} \caption{\small
Non-periodic first-order SUSY partner $\widetilde V^{np}(x)$ (black
curve) of the associated Lam\'e potential (gray curve) for $m=3, \
\ell=1$, $k^2=0.95$, which was generated by using $u(x) = \psi^+(x)
+ \psi^-(x)$ with $\epsilon = 4.75$. The new Hamiltonian $\widetilde
H^{np}$ has an extra bound state precisely at $\epsilon=4.75$.}
\end{figure}

Our first task is to evaluate four constants $a_1,a_2,a_3,a_4$ (note
that $m+\ell=4$) in (\ref{prs}), where without loss of generality,
we choose $a_0=1$. Let us first write down the basic elements $f_i$
for $m=3,\ell=1$ from (\ref{f0}-\ref{f2})
\begin{eqnarray}\label{s1}
& \hskip-0.5cm f_0(\rho)=\bar e_3\rho(\rho-3)(2\rho-3), \quad
 f_1(\rho)=2(\rho-1)\{3e_1[4-(\rho-1)^2]-\widetilde{E}\}, \\
\label{s2} & f_2(\rho)=\bar e_2(\rho-4)(\rho+3)(2\rho-1), \qquad
 \widetilde{E}=\bar e_3(E-2)+12e_3.
\end{eqnarray}
Note that the two constants $a_1,a_2$ have to be determined from
(\ref{a2l})
\begin{equation}\label{s3}
 a_1=-\frac{F_1}{f_0(1)}, \qquad a_2=\frac{F_2}{f_0(1)f_0(2)},
\end{equation}
while the remaining two are to be computed from (\ref{fa2l})
\begin{equation}\label{s4}
a_3=-\frac{D_1f_2(2)}{D_2}a_2, \qquad
a_4=\frac{D_0f_2(2)f_2(3)}{D_2}.
\end{equation}

Now from (\ref{Fr}), one may write quite straightforwardly the
$F_i$'s
\begin{equation}\label{s5}
 F_1=f_1(0), \qquad F_2=\left | \begin{array}{ll}
                                 f_1(1) & f_2(0) \\ f_0(1) & f_1(0)
                                 \end{array} \right |,
\end{equation}
and to obtain the $D_i$'s  we need
\begin{equation}
\label{F5} F_5=\left |
  \begin{array}{lllll}
     f_1(4)       &      f_2(3)        &   \hspace{.7cm} 0    &   \hspace{.7cm} 0   & \hspace{.7cm} 0 \\
     f_0(4)       &      f_1(3)        &       f_2(2)       &   \hspace{.7cm} 0   & \hspace{.7cm} 0 \\
  \hspace{.7cm} 0   &    f_0(3)        &       f_1(2)       &      f_2(1)       & \hspace{.7cm} 0 \\
   \hspace{.7cm} 0 & \hspace{.7cm} 0  &        f_0(2)       &      f_1(1)   &    f_2(0) \\
   \hspace{.7cm} 0 & \hspace{.7cm} 0  & \hspace{.7cm} 0     & f_0(1)   &    f_1(0) \\
      \end{array} \right |.
\end{equation}
Then the final quantities are
\begin{equation}\label{s7}
 D_0=[\mbox{minor of $F_5$ in $F_5$}]=1,
\end{equation}
\begin{equation}\label{s8}
D_1=[\mbox{minor of $F_4$ in $F_5$}]=f_1(4),
\end{equation}
\begin{equation}\label{s9}
 D_2=[\mbox{minor of $F_3$ in $F_5$}]=\left | \begin{array}{ll}
                                 f_1(4) & f_2(3) \\ f_0(4) & f_1(3)
                                 \end{array} \right |.
\end{equation}

Finally, using the ingredients (\ref{s1}-\ref{s9}) we obtain
\begin{eqnarray}
& a_0 = 1, \qquad a_1 = \frac{9e_1-\widetilde E}{\bar e_3}, \qquad
a_2 =
\frac{6\bar e_2}{\bar e_3}, \nonumber \\
& a_3 = -\frac{45(\widetilde E + 15e_1)\bar e_2^2}{(\widetilde E^2 +
15 e_1 \widetilde E + 25 \bar e_2\bar e_3)\bar e_3}, \quad a_4 =
\frac{225 \, \bar e_2^3}{(\widetilde E^2 + 15 e_1 \widetilde E + 25
\bar e_2\bar e_3)\bar e_3}.
\end{eqnarray}
We employ these coefficients then to find the roots $c_r,
r=1,\dots,4$ of the fourth-order equation
\begin{equation}
\sum_{r=0}^{4} a_r \left(\frac{e_1-t}{\bar e_2}\right)^r = 0,
\end{equation}
which can be analytically determined, but their explicit expression
is too involved to be shown here. These roots are used then to
invert the transcendental equation $\wp(b_r)= c_r$ to determine the
$b_r$'s (with the restriction $\Psi'\vert_{z=b_r}>0$), which are
thus inserted in the explicit expressions for $\psi^{\pm}(x)$.
Finally, the resulting Bloch solutions can be used, either directly
or in the corresponding Wronskian, to derive the periodic SUSY
partner potentials $\widetilde V_\pm(x)$ of (\ref{tvp1}) or
$\widetilde V(x)$ of (\ref{tvp}). On the other hand, different
linear combinations of kind (\ref{linearcombination}) or
(\ref{lc12}) can be used to derive the potentials $\widetilde
V^{np}(x)$ of (\ref{tvnp1}) or (\ref{tvnp2}) which have periodicity
defects. The final results of these procedures are illustrated in
figures below. In the four figures we show in gray the original
associated Lam\'e potential for $m=3$, $\ell = 1$, $k^2=0.95$. In
figure 1 we show as well in black one of its periodic first-order
SUSY partners generated through a Bloch solution with $\epsilon =
4.75$ while in Figure 2 it is illustrated one of its non-periodic
partners for the same $\epsilon$, which is lesser than but close to
the lowest band edge $E_0 = 4.79991$. On the other hand, in figures
3 and 4 we have drawn similar graphs (black curves) for the
corresponding second-order SUSY partners, periodic and non-periodic
respectively. For the periodic case (Figure 3) we have used two
Bloch solutions $u_1(x) = \psi_1^+(x)$, $u_2(x) = \psi_2^+(x)$
associated to the pair of factorization energies $\epsilon_1 = 9.4$,
and $\epsilon_2 = 9.5$ which are in the first finite energy gap
$(4.8,9.55)$. For the non-periodic case we have used the same pair
of factorization energies, with linear combinations $u_1(x) =
\psi_1^+(x) + \psi_1^-(x)$ and $u_2(x) = \psi_2^+(x)-2\psi_2^-(x)$.
In both non-periodic cases the periodicity defects are clearly
detected.

\begin{figure}[ht]
\centering \epsfig{file=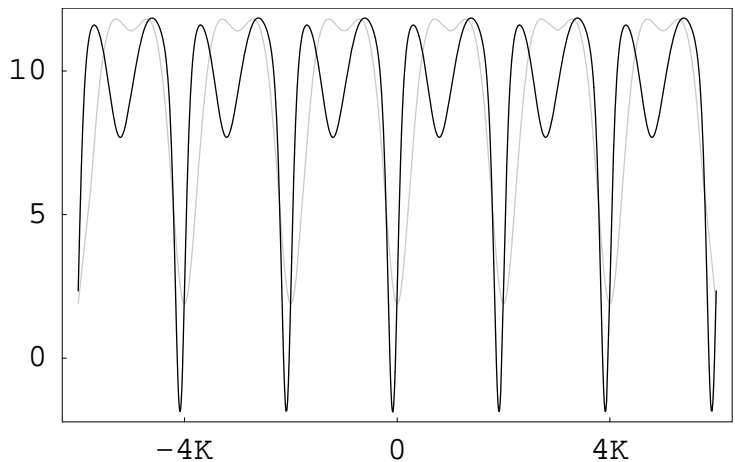, width=10cm} \caption{\small
Periodic second-order SUSY partner potential $\widetilde V(x)$
(black curve) isospectral to the associated Lam\'e potential (gray
curve) with $m=3, \ \ell=1$, $k^2=0.95$, generated by using
$\epsilon_1 = 9.4$, $u_1(x) = \psi_1^+(x)$ and $\epsilon_2 = 9.5$,
$u_2(x) = \psi_2^+(x)$.}
\end{figure}

\begin{figure}[ht]
\centering \epsfig{file=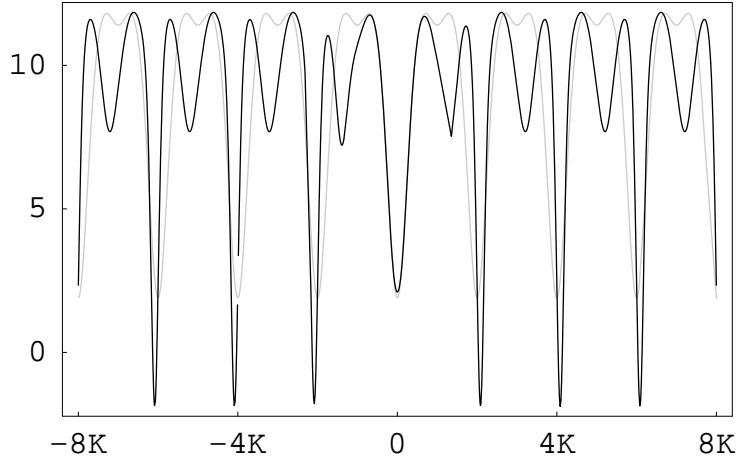, width=10cm} \caption{\small
Non-periodic second-order SUSY partner $\widetilde V^{np}(x)$ (black
curve) of the associated Lam\'e potential (gray curve) with $m=3, \
\ell=1$, $k^2=0.95$, which was generated by using $\epsilon_1 =
9.4$, $u_1(x) = \psi_1^+(x) + \psi_1^-(x)$ and $\epsilon_2 = 9.5$,
$u_2(x) = \psi_2^+(x)-2\psi_2^-(x)$. The Hamiltonian $\widetilde
H^{np}$ has the same band spectrum as the initial associated Lam\'e
potential plus two isolated bound states at $\epsilon_1$ and
$\epsilon_2$.}
\end{figure}

\section{Conclusions}

In this article we have shown, by finding explicitly the two
solutions of Bloch type associated to the stationary Schr\"odinger
equation for an arbitrary value of the energy, that the associated
Lam\'e potentials for any integer values of the parameter pairs
$(m,\ell)$ are exactly solvable. This point is clarifying because in
most of the works concerning Schr\"odinger equation with periodic
potentials typically one looks for just the band edge
eigenfunctions, thus inducing the idea that these solutions are the
only ones which could be analytically determined. The solved problem
is very important for implementing the supersymmetric
transformations, of first or higher order, since the non-physical
Schr\"odinger solutions can be used as seeds to generate new exactly
solvable potentials. Consequently, we have generated in this way
potentials which are either strictly isospectral to the initial one
or with some isolated bound states embedded at the energy gaps of
the initial Hamiltonian. The arising of these two different kinds of
spectra for the new potentials depends on either we choose as seeds
directly the Bloch solutions or general linear combinations for the
chosen factorization energies. The potentials with bound states
embedded into the finite gaps may be interesting for physical
applications since the corresponding levels could work as auxiliary
transition energies for the electron to jump easier between allowed
bands.

\section*{Acknowledgment}
This work was partially supported by the post-doctoral grant of AG
(No. 0000147287) from the Spanish Ministry of Foreign Affairs. DJFC
acknowledges the support of Conacyt (project No. 49253). The authors
acknowledge the warm hospitality at Department of Theoretical
Physics, Atomic and Optics, University of Valladolid, Spain, where
this work was finished. AG acknowledges City College authorities for
study leave.

\newpage

\appendix
\section{Appendix}
In the following we will provide a short introduction about elliptic
functions (for more details, see \cite{gr,as}). Three Jacobian
elliptic functions are defined by
\begin{equation}\label{adef1}
{\rm sn}(x,k)=\sin \varphi, \quad {\rm cn}(x,k)=\cos\varphi, \quad
{\rm dn}(x,k)=d\varphi/dx,
\end{equation}
where amplitude function $\varphi (z,k)$ is defined by the integral
\begin{equation}\label{adef2}
z(\varphi,k)=\int_0^\varphi\frac{d\tau}{\sqrt{1-k^2\sin^2\tau}}.
\end{equation}
The square of the real number $k$ is called elliptic modulus
parameter and $k^2\in(0,1)$. $k'^2=1-k^2$ is called complementary
modulus parameter. For simplicity, in the text we suppress the
explicit modular dependence and write ${\rm sn}\,x,\ {\rm cn}\,x,\
{\rm dn}\,x$, etc. These are doubly periodic functions of periods
$4K,2iK'$; $4K,4iK'$ and $2K,4iK'$ respectively, where the
quarter-periods $K$ and $K'$ are the real numbers given by
\begin{equation}\label{adef3}
 K(k)\equiv K=z(\pi/2,k), \qquad K'(k)\equiv K'=K(k').
\end{equation}
$K$ is called complete elliptic integral of second kind. Some
relevant relations are
\begin{equation}\label{adef4}
{\rm sn}(x+K)=\frac{{\rm cn} \, x}{{\rm dn} \, x}, \quad {\rm
cn}(x+K)=-k'\frac{{\rm sn} \, x}{{\rm dn} \, x}, \quad {\rm
dn}(x+K)=\frac{k'}{{\rm dn} \, x},
\end{equation}
\begin{equation}\label{adef5}
{\rm sn}(x+2K)= -{\rm sn} \, x, \quad {\rm cn}(x+2K)=-{\rm cn} \,
x,\quad {\rm dn}(x+2K)={\rm dn} \, x,
\end{equation}
\begin{equation}\label{adef6}
{\rm sn}(x+iK')=\frac{1}{k{\rm sn} \, x}, \quad {\rm
cn}(x+iK')=-\frac{i}{k}\frac{{\rm dn} \, x}{{\rm sn} \, x}, \quad
{\rm dn}(x+iK')=-i\frac{{\rm cn} \, x}{{\rm sn} \, x},
\end{equation}
\begin{equation}\label{adef7}
{\rm sn}^2x+{\rm cn}^2x = 1, \qquad {\rm dn}^2x+k^2{\rm sn}^2x=1,
\end{equation}
and the rules of differentiation are
\begin{equation}\label{adef8}
{\rm sn}'x={\rm cn}\, x \ {\rm dn}\, x, \quad {\rm cn}'x=-{\rm sn}\,
x \ {\rm dn}\, x, \quad {\rm dn}'x=-k^2{\rm sn}\, x \ {\rm cn} \, x.
\end{equation}

Weierstrass elliptic function $\wp (z;g_2,g_3)\equiv \wp (z)$ is
defined by
\begin{equation}\label{p1}
\wp (z)=\frac{1}{z^2}+\sum_{m,n}\hspace{-1pt}^{'}
 \left[\frac{1}{(z-2m\omega-2n\omega')^2}-\frac{1}{(2m\omega+2n\omega')^2}\right],
 \end{equation}
where the symbol $\sum'$ means summation over all integer values of
$m,n$ except $m=n=0$; $\omega,\ \omega'$ being half-periods of $\wp
(z)$. The invariants $g_2,g_3$ are
 given by
\begin{equation}\label{p2}
 g_2=60\sum_{m,n}\hspace{-1pt}^{'} \frac{1}{(m\omega+n\omega')^4}, \qquad
    g_3=140\sum_{m,n}\hspace{-1pt}^{'}
    \frac{1}{(m\omega+n\omega')^6}\, .
 \end{equation}
 The three numbers $e_i,i=1,2,3$ are defined by
 $\wp(\omega_i)=e_i$, where
 $\omega_1\equiv\omega, \ \omega_3\equiv\omega'$ and
 $\omega_2=\omega+\omega'$.
To get the relation between Jacobian elliptic functions and
Weierstrass elliptic function, it is necessary to define
$\omega,\omega'$ in terms of $K,K'$. We have taken the following
definition
\begin{equation}\label{om}
 \omega=\frac{K}{\sqrt{\bar e_3}}, \qquad \omega'=\frac{i K'}{\sqrt{\bar e_3}},
\end{equation}
which corresponds to the case when the discriminant
$\Delta=g_2^3-27g_3^2>0$. This means that the numbers $e_i$ are
always real and consequently they can be ordered as $e_1>e_2>e_3$,
because these are roots of the equation
\begin{equation}\label{e1}
4t^3-g_2t-g_3=0.
\end{equation}

The relations between $\wp(z)$ and ${\rm sn}\, z,\ {\rm cn}\, z,\
{\rm dn}\,z$ may then be written as
\begin{equation}\label{r1}
 \wp\left(\frac{z}{\sqrt{\bar{e}_3}}\right)=
 e_1+\bar{e}_3\frac{{\rm cn}^2z}{{\rm sn}^2z}=e_2+\bar{e}_3\frac{{\rm dn}^2z}{{\rm sn}^2z}
          =e_3+\bar{e}_3\frac{1}{{\rm sn}^2z}.
\end{equation}
It may be mentioned that the derivative of $\wp(z)$ is also an
elliptic function with same periods and satisfy the relation
\begin{equation}\label{p3}
\wp'^2(z)=4\:\prod^3_{i=1}\left[\wp (z)-e_i\right].
\end{equation}

It is now straightforward to obtain equation (\ref{aw}) from
equation (\ref{aj}) under the translation $x=\sqrt{\bar{e}_3}z+iK'$
by using the relations (\ref{adef6}) and (\ref{r1}). We will now
apply the transformation $y=(e_1-\wp(z))/\bar{e}_2$
 on equation (\ref{prod1}). Noting the following relations
\begin{equation}\label{p4}
 \wp''(z)=6\wp^2(z)-\frac{g_2}{2}, \qquad
 \frac{\wp'''(z)}{\wp'(z)}=12\wp(z),
\end{equation}
 it is not very difficult to obtain the following equation
\begin{eqnarray}
&& 2\prod^{3}_{i=1}\big[\wp(z)-e_i\big]\frac{d^3\Psi}{dy^3} -
3\bigg[3\wp^2(z)+\sum^{3}_{\renewcommand{\arraystretch}{.3}
                       \begin{array}{c}{\scriptstyle i,j=1} \\
                                    {\scriptstyle i\neq j}
                       \end{array}} \renewcommand{\arraystretch}{1}e_ie_j\bigg]\bar{e}_2\frac{d^2\Psi}{dy^2}
\nonumber  \\ && + 2\bigg\{[3-m(m+1)]\wp(z)
-\frac{\ell(\ell+1)\bar{e}_2\bar{e}_3}{\wp(z)-e_1}+
\widetilde{E}\bigg\}(\bar{e}_2)^2\frac{d\Psi}{dy} \nonumber \\
&&
+\bigg\{m(m+1)-\frac{\ell(\ell+1)\bar{e}_2\bar{e}_3}{[\wp(z)-e_1]^2}\bigg\}(\bar{e}_2)^3\Psi=0
\label{int1}
\end{eqnarray}
The equation (\ref{prod2}) will then readily follow under the
transformation $\Phi(y)=[\wp(z)-e_1]^{\ell}\Psi$ on above equation
by using the relations (\ref{p4}).

\end{document}